# Virtual Guide Dog: Next Generation Pedestrian Signal for the Visually Impaired

Zijia Zhong[1] and Joyoung Lee[2]

## Abstract

Accessible pedestrian signal (APS) was proposed as a mean to achieve the same level of service that is set forth by the American with Disability Act (ADA) for the visually impaired. One of the major issues of existing APSs is the failure to deliver adequate crossing information for the visually impaired. This paper presents a mobile-based APS application, namely Virtual Guide Dog (VGD). Integrating intersection information and onboard sensors (e.g., GPS, compass, accelerometer, and gyroscope sensor) of modern smartphones, the VGD application can notify the visually impaired: 1) the close proximity of an intersection and 2) the street information for crossing. By employing a screen tapping interface, VGD can remotely place a pedestrian crossing call to the controller, without the need of using a push button. In addition, VGD informs VIs the start of a crossing phase by using text-to-speech technology. The proof-of-concept test shows that VGD keeps the users informed about the remaining distance as their approaching the intersection. It was also found that the GPS-only mode is accompanied by greater distance deviation compared to the mode jointly operating with both GPS and cellular positioning.

## Introduction

The American with Disability Act (ADA) requires the same level of service to be provided to the visually impaired as to others.[1] Despite the significant improvements for accessible pedestrian signal (APS) systems over the years, APSs are still facing many challenges. For example, the visually impaired has to deviate from their travel path in order to reach the push button to actuate the pedestrian phase. A study showed that less than 16% of the participating the visually impaired persons who looked for a push button and successfully found one.[2] The APS also brings externality to the surrounding environment: the repeating tone from APS adds 5 decibels of noise within a 12-ft. radius. Besides, the installation cost for each APS system is approximately $6000 per intersection with labor cost included.[3]

Even with APSs, the the visually impaired is still experiencing difficulties in crossing signalized intersection due to the lack of safe crossing information. Research showed that only 49% of them started crossing during the walk interval and 27% of all independent crossings ended after the onset of the conflicting traffic.[2] It is more the case when an available push button is not used: the walk time is shorter than necessary for crossing under such circumstance. Admittedly, there are several state-of-the-art solutions that have been designed specifically for the the visually impaired, including electronic white cane[4] with integrated camera[5], wearable belt with GPS and compass[6], BrailleNote GPS[7], radio frequency identification (RFID), and ultrasonic-to-audio guidance system[8]. However, the majority of these devices has limited market penetration and they are still inadequate in one of the crucial aspects of a signalized intersection, which is the communications between the users and the signal controllers. Owing to the advancement of the information and communication technologies, a new generation of APS with a user-friendly interface can be developed. This study proposes a mobile-based APS, namely, the Virtual Guide Dog (VGD), which provides the visually impaired users with personalized intersection crossing instructions and improves the safety of the crossing. Not only does VGD provide a more interactive and yet higher usable application to the the visually impaired, but it is also able to make the full use of the existing "infrastructure", as smartphones have evolved to an integral part of the modern society.

The remainder of the paper is organized as follows. In the "Literature Review" section, the challenges of APS for the the visually impaired as well as relevant research efforts are summarized. Then the proposed application architecture, test bed, and application logic are presented in the "System Architecture" section. Results of the preliminary POC field test are discussed in the "Evaluation" section, followed by the "Conclusion" section.

[1] Department of Civil and Environmental Engineering, University of Delaware, United States.
[2] John A. Reif Jr. Department of Civil and Environmental Engineering, New Jersey Institute of Technology, United States

**Corresponding author:**
Joyoung Lee, John A. Reif Jr. Department of Civil and Environmental Engineering, New Jersey Institute of Technology, Newark, NJ 07102, USA

Email: jo.y.lee@njit.edu

## Literature Review

Accurate construction of a representation of the world is challenging for the person with loss of vision[9]. Besides physical inability, the "inaccessibility" was also characterized by the difficulty in acquiring information and spatial knowledge[10]. The visually impaired primarily relies on the auditory, olfactory, and somatosensory (haptic and temperature) cues to acquire the aforementioned information for navigation.[11]

Wayfinding systems have been developing to aid the visually impaired to travel independently. Rodrigues-Sanchez et al.[12,13] proposed an accessible wayfining service system that focuses on people with a disability in both outdoor and indoor environments. Apostolopoulos et al.[14] developed an online localization system which helped the visually impaired navigate in an indoor environment by utilizing inexpensive sensors which were readily available in current smartphones. The system tracked the location of the user by sensor readings, knowledge of the indoor environment (e.g., architectural blueprints), and possible landmark confirmed by users. The system employed text-to-speech technology for interfacing with users. One of the drawbacks for the system was the inaccurate localization which was caused by electromagnetic noise in an indoor environment and mediocre quality of the built-in sensors in certain smartphones.

Crossing an intersection introduces an extra layer of complexity to the typical wayfinding for the visually impaired. The visually impaired is more vulnerable to collision, especially when crossing an intersection, due to insufficient information concerning traffic, walk phase, and intersection geometry.[7] A seemingly effortless crossing at intersection for sighted pedestrians has been proven considerably difficult for the visually impaired. The sub-tasks involving in crossing a street can be broken down to

1) detecting the existence of an intersection,
2) determining the street or direction (heading) to cross,
3) locating the crosswalk,
4) aligning the heading with the destination, and
5) maintaining a straight path during crossing within a crosswalk.

The visually impaired mainly relies on the sound created by the parallel movement of traffic as cues for starting the crossing at signal crossings that are not equipped with APSs[15]. With the presence of APSs, the beeps and chirps from APS provided directional crossing cues. However, the persistent noises from the street can mask the target signal (APS) and consequentially affect orientation and mobility for directional travel for the visually impaired.[11]

Furthermore, intersection geometric features (e.g., dedicated right-turning lanes, multiple lanes, rounded corners), which are designed to move vehicles as efficiently as possible, inevitably increase the difficulties for the crossing of the visually impaired. The National Cooperative Highway Research Program (NCHRP) conducted an extensive study[16] regarding the accessibility and crossing solutions at roundabouts and channelized turn lanes for pedestrians with vision disability. It was found that the surveyed intersections failed to provide reliable information for the following principle tasks, which are 1) determining crosswalk location, 2) aligning to cross, and 3) maintaining heading during crossing.

Not until the year of 2000 was audible pedestrian signals included in the Manual on Uniform Traffic Control Devices (MUTCD)[17]. Chapter 4E in the MUTCD regulates the pedestrian control feature, including APSs and detectors. The manual specifies the standards in terms of location, walk indication, tactile arrows and locator tones, and extended push button press feature from Section 4E.09 to 4E.13. Under the standards set forth by the MUTCD, the required features for an APS system include:

1) tone for the push button location
2) vibrotactile arrow indicating the crossing direction on the push button
3) vibrotactile indication of beginning of the "walk" phase
4) audible "walk" phase indication tone (e.g., percussive tone or speech walk message)
5) automatic volume adjustment function which maintains volume above the ambient sound level

Guth et al.[18] developed an intersection database that provided intersection and crossing information to the visually impaired via a braille-based personal data assistant (PDA) device. Four categories of intersection information were available: 1) intersection shape and size, 2) crosswalks and curb ramps, 3) traffic signals and control, and 4) accessible pedestrians signals. In the field experiments, they found that significant positive effects for database description on crossing subtasks, including crossing initiation, use of the pushbutton. The positive effect, however, was not shown in the other subtasks, such as maintaining a straight path in a crosswalk. They also found that the likelihood of completing crossing was nine times higher in the presence of APS. With APS, a higher percentage of crossing completion before the end of crossing signal was observed.[19]

Bentzen et al.[20] conducted a before-and-after study to examine the impact of APS equipped with beacon features in two U.S. cites. The performance of the APS was evaluated based on two main categories: timing measure and wayfinding measure. Results showed that numerous improvements were achieved, including reduced stating delay, higher rates in finding starting location, and better crosswalk alignment. Despite the improvements thus far, APS still faces challenges. First of all, the current APS requires users to know their travel direction in advance. Second, the actuation of pedestrian phase (via a push button) requires the visually impaired pedestrians deviate from their path of travel in most instances. Third, its above-ambient-volume requirement increases 5 decibels of noise on average in the surrounding area; lastly, on-going maintenance for the APS is necessary and could be costly.

Wilson et al.[21] developed an assistance device for the visually impaired, called System for Wearable Audio Navigation (SWAN). It addressed the limitations of previous speech-based navigation aids by adapting non-speech audio presentation of navigation information when possible. It also enabled the ability to track users' current location and heading. Additionally, it provided guidance for a near optimal and safe walking path to destinations, and the awareness of salient features of the path.

Ramadhan et al.[22] proposed a wearable smart system that integrated a microcontroller (Arduino Uno), cellular communication, GPS module, and a solar panel. Three types of sensors (i.e. ultrasonic sensor, accelerometer, and voice recognition sensor) were used to track the path and alert the users of obstacles. The system was also equipped with remote monitoring capability which could be made available to family and caregivers. Lan et al.[23] developed a smart glass system whose main function was to see the world for the visually impaired, with a focus on public sign recognition. A mobile-based personal APS, named mobile accessible pedestrian signals (MAPS), was proposed by Liao[24]. The main function of the MAPS system was to provide the visually impaired with available intersection geometry condition as well as signal timing information through a smartphone application. By using built-in sensors of a smartphone (e.g., GPS and digital compass) along with signal phasing and timing plans, not only can the MAPS inform pedestrian when to cross, but also how to align with the crosswalk. Since the signal plans are centrally stored within the Minnesota Department of Transportation, necessary cyber-security infrastructures had to be placed including virtual private network tunneling, firewall, and request authentication.

The VGD application was motivated by the fact that the majority of the wayfinding applications were passive systems, which mean there was little, if any at all, communication between the visually impaired and the traffic signal controllers. In this paper, we proposed a pedestrian crossing mobile application which could send crossing requests to the signal controller via The National Transportation Communication for Intelligent Transportation System Protocol (NTCIP) protocol without the need for pushing the conventional actuation push button. The NTCIP[25] is a family of standards intended for use in all types of management systems concerning transportation environment (e.g., traffic signals, transit, emergency management, data archiving). The NTCIP was developed to provide communication standards which ensure the interoperability and interchangeability among traffic controllers and ITS devices. Owing to NTCIP, a user-orientated and personalized APS could be developed.

## System Architecture

The high-level architecture for VGD system is shown in Figure 1. The VGD mobile application continuously accesses the built-in GPS and compass units to determine the location and heading of a user, respectively. When a user reaches the proximity of an intersection, the VGD informs the user the proximity of an intersection and ask the user about the direction or street he or she wants to go. When a user arrives at the intersection, the application retrieves available crossing information (e.g., direction, street names) for the user. A dedicated database containing the geometry information of intersections can be established in advance.

Alternatively, open source GIS (e.g., Google Maps API[26]) could also be utilized. Once the user has selected the path and stated the street name of choice, the VGD application verbally instructs user to turn to the heading of crossing and, at the same time, sends the crossing request to the controller to actuate the pedestrian crossing phase. Once the crossing phase starts, the application informs the user that it is safe to cross the roadway and the remaining time for the green phase.

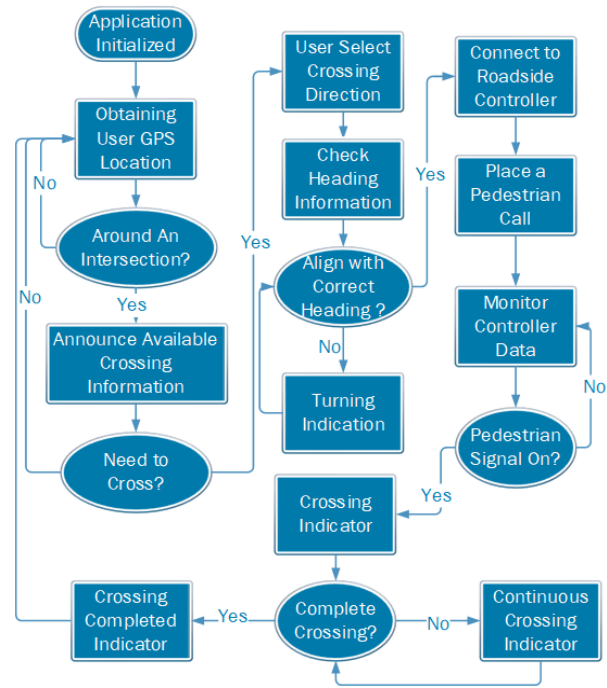

**Figure 2.** VGD mobile application logic

Figure 2 shows an overall procedure implemented by the VGD mobile application. The VGD mobile application plays a crucial role in enabling the VGD system to communicate with the users of the information that is necessary for safe and prompt crossing. The primary functions of the applications include:

- constant GPS-based localization for a pedestrian on a roadway and crossing information based on proximity to the intersection

- touchable and audible user interface for users to exchange information

- pedestrian phase actuation without the need to press the push button

- instructions, if necessary, for pedestrians to ensure alignment with the crosswalk

- wireless communication with a traffic signal controller via Bluetooth

Currently, the VGD application is available on any Android-based mobile devices. Figure 3 shows the main screen of the VGD application running on an Android smartphone. Note that the VGD application is fully expected to make available on iOS-based devices (e.g., iPhone, iPad)

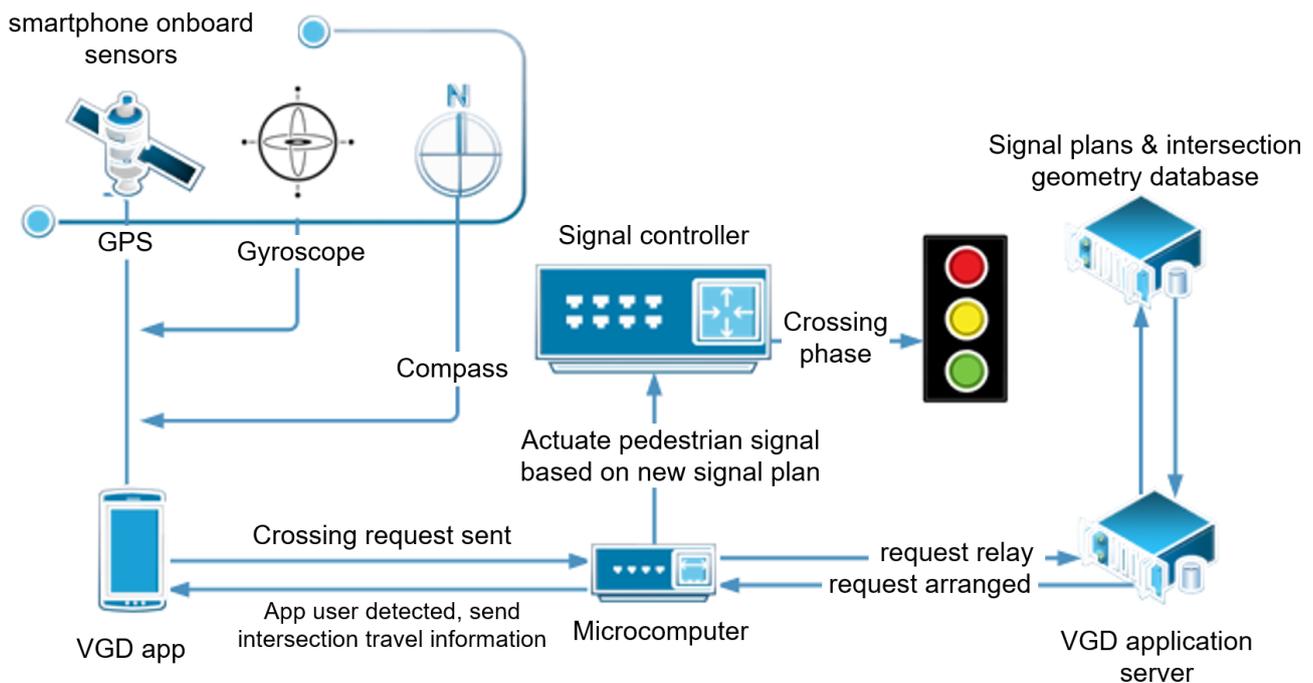

**Figure 1.** High-level architecture for VGD

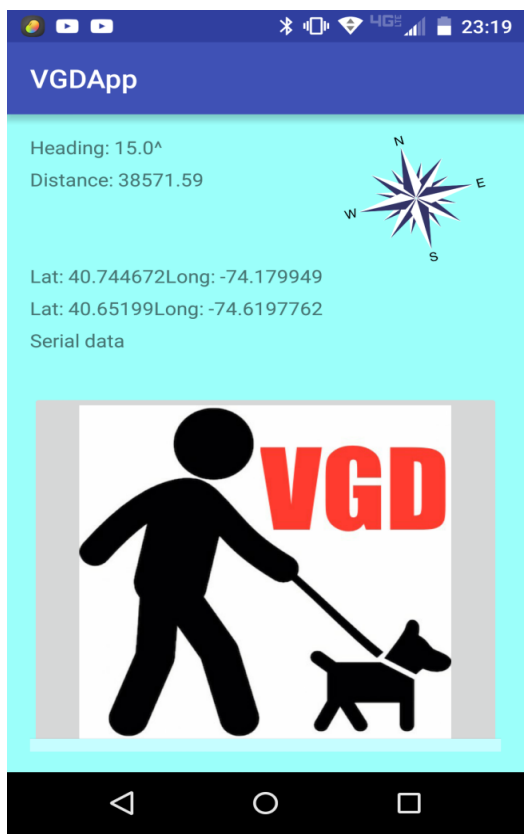

**Figure 3.** VGD Andriod application interface

## Proof-of-Concept Test

The use of VGD enables the visually impaired to save certain steps which would have been necessary. The steps that could be saved by using VGD application include: 1) determination if an intersection is within the proximity 2) determination of the location of a crossing push button, if exists, 3) the excursion to the push button from the path of crossing 4) the determination of the start of a crossing phase All of the steps saved above potentially results in a short crossing cycle (i.e., from the beginning of the crossing need the completion of crossing an intersection) by providing additional information (e.g., street name and heading information) to reduce users' uncertainties. A shorter crossing time also enhance users' satisfaction.

### Simulation-based Test

The application was first tested in a hardware-in-the-loop simulation test bed. The primary purpose of the simulation was to test whether the application could successfully place crossing request at the controller by using NTCIP protocol. As shown in Figure 4, the proposed HILS test bed is comprised of 1) Vissim (a microscopic traffic simulator), 2) Vissim Component Object Model (COM) Interface 3) a hardware signal controller, and 4) the VGD application An production signal controller was used to test potential integration issues. A simulation manager program written in Vissim COM interface is used to coordinate all the simulation components. During the simulation, pedestrian crossing requests were generated and sent to signal controller for actuation. The Vissim COM Interface read the real-time signal data via NTCIP and update the Vissim simulation. Simulation test showed that the VGD can successfully place requests to the controller.

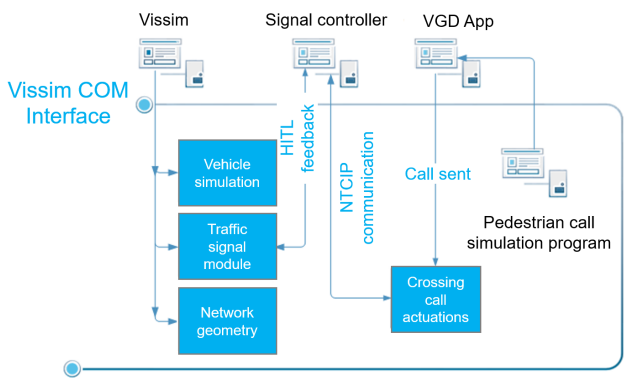

**Figure 4.** Hardware-in-the-loop simulation framework

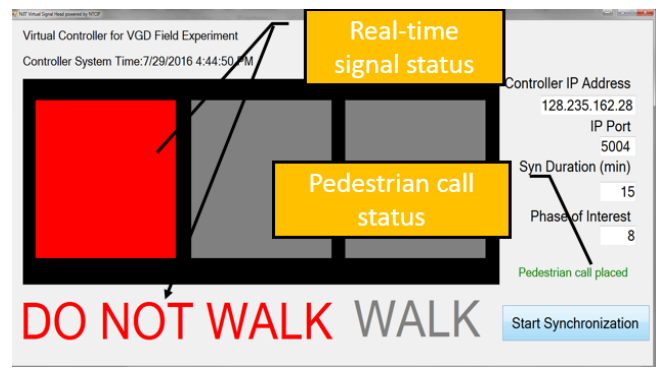

**Figure 6.** Virtual controller signal head

## Field Test

The second stage of the test is to examine the VGD application in the field. An intersection around the New Jersey Institute of Technology (NJIT) was selected for the field demonstration. Located at the intersection of Central Ave. and Lock St. in Newark, NJ, the selected intersection has a high pedestrian crossing demand due to its proximity to the university campus. A microcomputer which was capable of communicating with the mobile application and relaying crossing request remotely was added.

For the field evaluation, an Econolite Cobalt ATC controller situated in the Intelligent Transportation System Laboratory (ITSL) at NJIT, which was retrofitted with a microcomputer and Bluetooth scanner as shown in Figure 5, was used to test the controller integration. The pedestrian crossing calls were placed remotely from the field intersection via NJIT campus wireless network or 4G/LTE cellular network. In order to monitor the real-time status on the controller remote location, as one would see the overhead signal, a virtual signal program which independently synchronizes and displays the controller signal head status (6) was developed. The ground station for the field test is shown in Figure 7.

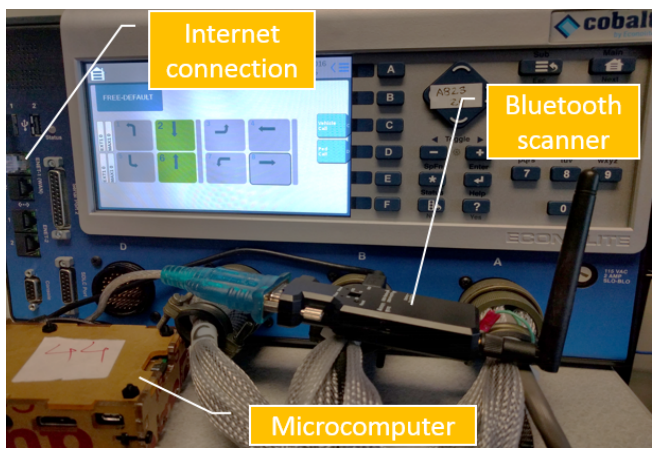

**Figure 5.** Controller retrofitted with microcomputer and Bluetooth scanner

During the test, volunteer users were assigned with Android smartphones installed with the VGD application. They were instructed to walk to the intersection as shown in Figure 8. After being initialized, the application continuously updates a users location and the distance to the intersection. If the distance is greater than 500 m, the application does not produce any verbal notification. When the user reaches within 500 m radius of an intersection, the application notifies the user of the distance in every 100 m. Upon arrival of the intersection, the application announces the street and heading information. User can loop through the available crossing directions by a short tapping on the smartphone screen. The crossing call is placed by a longer tapping on the screen after the intended crossing direction is announced. The evaluation results showed that the application is able to remotely communicate with the controller via NTCIP protocol and place crossing requests successfully. More importantly, the application can reliability actuate the controller pedestrian phase. Only the Cobalt ATC controller was used during the test, but it is expected that the VGD can be easily integrated with other signal controllers which are NTCIP compatible.

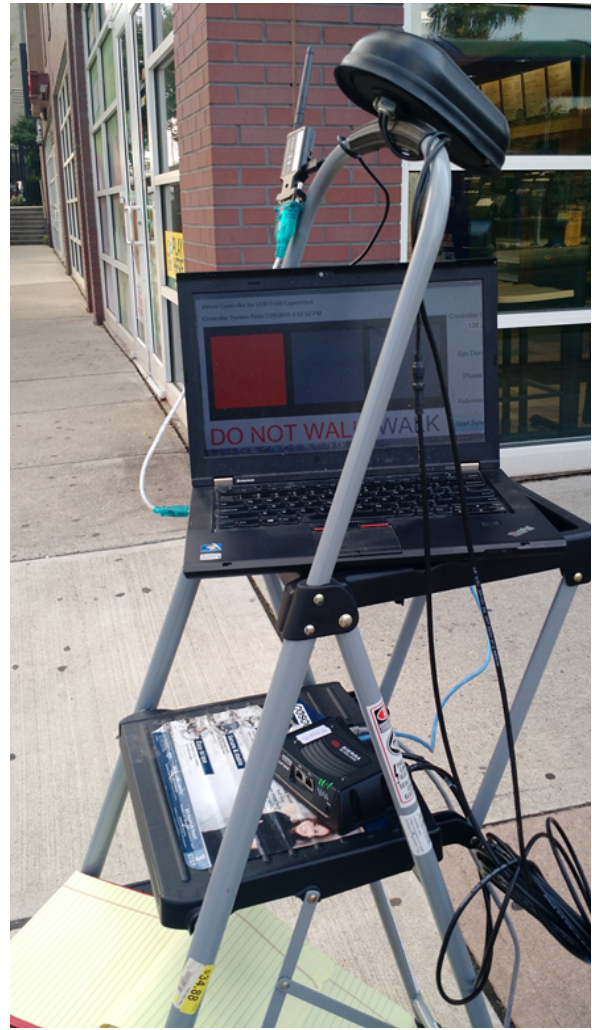

**Figure 7.** Field test setup

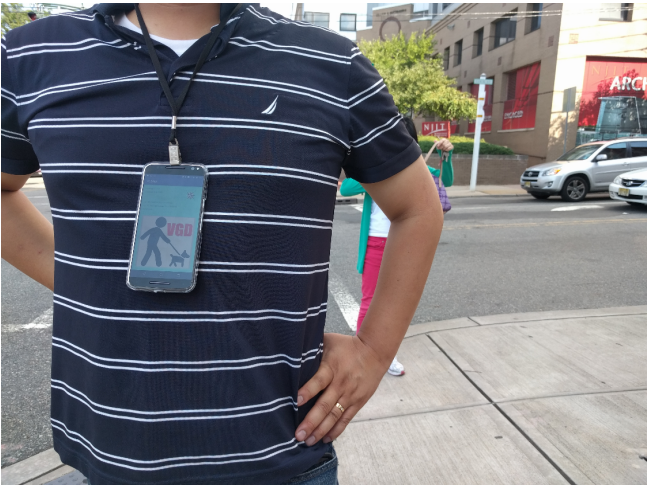

**Figure 8.** Field test participant

*Results and Findings*

The accuracy of position based on either onboard GPS or cellular-enhanced GPS was tested and the results are shown in Figure 9. The distance to the intersection from the starting point was 58.5 m away from the intersection and four intermediate reference points were recorded separately during the walk. The application showed 18-m closer than ground truth for enhanced-GPS mode, whereas the distance was measured 14.8 m closer to the intersection than ground truth for the GPS mode. The distance deviations exhibited a decreasing trend as the user getting closer to the intersection. A 1.7-m closer (5%) distance was recorded for enhanced GPS mode at the reference point #2 and 2.9-m closer (8%) of distance deviation was measured at the reference point #3. For the GPS only mode, fluctuations was more obvious than cellular-enhanced mode: the deviation was ranging between negative and positive values among reference points, for instance, the deviation changed from 5% further at the reference point #2 to 12% closer at the reference point #3. When users reached close proximity of the intersection (reference point #4), the distance deviation increased slightly for both modes. Overall, the distance measured by the enhanced GPS mode is more consistent than that of GPS mode

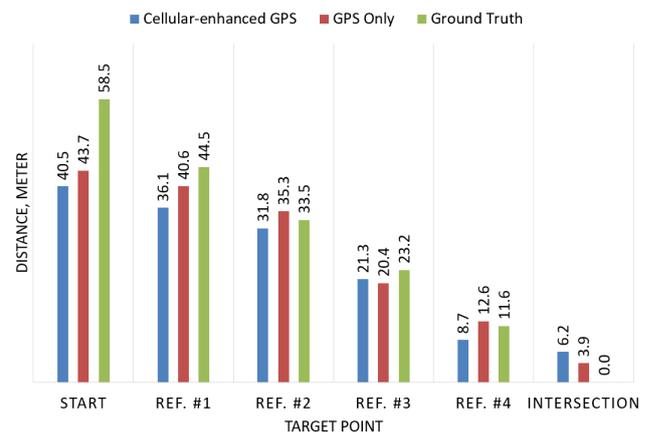

**Figure 9.** Comparison of distance information

The field test demonstrated that the VGD application was able to inform users the proximity of an intersection and provide intersection. Crossing requests were promptly relayed by the microcomputer and placed via NTCIP protocol. However, there were some issues discovered during the field experiment.

- Sometimes the built-in GPS experienced difficulty to pinpoint the location of the users. It could be attributed to the quality of the built-in GPS transponder as well as the default GPS filtering algorithm.

- As experienced in all GPS-based applications, the environment could impact the accuracy of the GPS. For instance, the GPS signal strength is weaker during cloudy days or under the shade of trees.

- Compass deviation was observed when an unaware user tilted the smartphone at an angle which exceeded certain threshold.

## Conclusions

In this study, we proposed a mobile Virtual Guide Dog application for the visually impaired users' safe intersection crossing. Using NTCIP protocol to communicate with the signal controller, the VGD application could be an attractive alternative for conventional APSs. The cost of implementing VGD is only a fraction of the conventional APS: average cost including labor for installing one APS system for one intersection is of \$6,000[3]; whereas the estimated cost for implement VGD is approximately from \$100 to \$500 depending on the price of mobile device and data plan. A remote signal controller was adopted for the field test, which demonstrated the success of using the NTCIP protocol, in conjunction with the VGD application, to make crossing request for the visually impaired persons using their smartphones.

The field test revealed possible future improvements as well. First, more signal controllers of different vendors should be tested for VGD interoperability. In addition, approaches to make VGD work with non-NTCIP-compliant signal controllers is needed. In such extreme cases, additional software/equipment may be needed for integrating the VGD system into incompatible signal controllers. Second, more advanced GPS filtering algorithms that are specifically designed for low-speed (i.e. walking speed) condition could be adapted to enhance the positioning accuracy. For example, the pedestrian dead reckoning is an effective way to obtain pedestrians location by estimated distance traveled via inertial sensors[27] and it can be used as a supplement for localization. Lastly, feedbacks from the visually impaired users are desired for further improvements. Focus groups of experienced and inexperienced eligible users could be conducted to further investigate the user preferences.

## Acknowledgements

This work is supported in part by the National Science Foundation under Grant No. CMMI-1844238